\documentclass[pre,floatfix,twocolumn,showpacs,a4paper,twoside]{revtex4}

\usepackage{psfig}
\usepackage{graphicx}
\usepackage{amsfonts,amssymb}
\usepackage{times}

\renewcommand{\vec}[1]{\mbox{\boldmath $#1$}}
\def\Om{\Omega}
\def\q{\qquad}
\def\beg{\begin{eqnarray}}
\def\ende{\end{eqnarray}}
\def\gsim{\lower.4ex\hbox{$\;\buildrel >\over{\scriptstyle\sim}\;$}} 
\def\lsim{\lower.4ex\hbox{$\;\buildrel <\over{\scriptstyle\sim}\;$}}

\renewcommand{\textrm} [1] {\rm #1} 
\renewcommand{\textit} [1] {\it #1}
\renewcommand{\textbf} [1] {\bf #1} 

\begin{document}

\title{Dissipative Taylor-Couette flows under the influence of
       helical magnetic fields}
\author{G. R\"udiger, M. Gellert, M. Schultz}
\affiliation{Astrophysikalisches Institut Potsdam,
         An der Sternwarte 16, D-14482 Potsdam, Germany}
\email{gruediger@aip.de, mgellert@aip.de,  mschultz@aip.de}
\author{R. Hollerbach}      
\affiliation{Department of Applied Mathematics, University of 
Leeds, Leeds, LS2 9JT, UK}
\email{rh@maths.leeds.ac.uk}

%\date{Accepted  Received  ;}
\date{\today}

\begin{abstract}
The linear stability of MHD Taylor-Couette flows in axially unbounded
cylinders is considered, for magnetic Prandtl number unity. Magnetic
fields varying from purely axial to purely azimuthal are imposed, with
a general helical field parameterized by $\beta=B_\phi/B_z$. We map out
the transition from the standard MRI for $\beta=0$ to the nonaxisymmetric
Azimuthal MagnetoRotational Instability (AMRI) for $\beta\to \infty$. For
finite $\beta$, positive and negative wave numbers $m$, corresponding to
right and left spirals, are no longer identical. The transition from
$\beta=0$ to $\beta\to\infty$ includes all the possible forms of MRI with
axisymmetric and nonaxisymmetric modes. For the nonaxisymmetric modes,
the most unstable mode spirals in the opposite direction to the
background field. The standard ($\beta=0$) MRI is axisymmetric for weak
fields (including the instability with the lowest Reynolds number) but is
nonaxisymmetric for stronger fields.

If the azimuthal field is due in part to an axial current flowing
through the fluid itself (and not just along the central axis), then it
is also unstable to the nonaxisymmetric Tayler instability, which is
most effective without rotation. For large $\beta$ this instability has
wavenumber $m=1$, whereas for $\beta\simeq 1$ $m=2$ is most unstable.
The most unstable mode spirals in the same direction as the background
field.
\end{abstract}


\maketitle
%%%%%%%%%%%%%%%%%%%%%%%%%%%%%%%%%%%%%%%%%%%%%%%%%%%%%%%%%%%%%%%%%%%%%%
\section{Introduction}
%%%%%%%%%%%%%%%%%%%%%%%%%%%%%%%%%%%%%%%%%%%%%%%%%%%%%%%%%%%%%%%%%%%%%%%
The longstanding problem of the generation of turbulence in various
hydrodynamically stable situations has found a solution in recent years
with the MHD shear flow instability, the so-called magnetorotational
instability (MRI), in which the presence of a magnetic field has a
destabilizing effect on a differentially rotating flow with outward
decreasing angular velocity but increasing angular momentum
\cite{BH91,V59}.

In the absence of MHD effects, according to the Rayleigh criterion, an
ideal flow is stable against axisymmetric perturbations whenever the
specific angular momentum increases outward
\beg
\frac{{\rm{d}}}{{\rm{d}}R}(R^2\Omega)^2 > 0,
\label{ray}
\ende
where $\Omega$ is the angular velocity, and ($R$, $\phi$, $z$) are
cylindrical coordinates.  In the presence of an azimuthal magnetic
field $B_\phi$, this criterion is modified to
\beg
\frac{1}{R^3}\frac{{\rm{d}}}{{\rm{d}}R}(R^2\Omega)^2-\frac{R}{\mu_0\rho}
\frac{{\rm{d}}}{{\rm{d}}R}\left( \frac{B_\phi}{R} \right)^2 > 0,
\label{mich}
\ende
where $\mu_0$ is the permeability and $\rho$ the density \cite{M54}.
Note also that this criterion is both necessary and sufficient for
(axisymmetric) stability. In particular, {\em all} ideal flows can thus
be destabilized, by azimuthal magnetic fields with the right profiles
and amplitudes.  

On the other hand, for nonaxisymmetric modes, one has
\beg
\frac{{\rm{d}}}{{\rm{d}}R}( R B_\phi^2) < 0
\label{tay}
\ende
as the necessary and sufficient condition for stability of an ideal
fluid at rest \cite{T73}.  Outwardly increasing fields are therefore
unstable. If  (\ref{tay}) is violated, the most unstable mode has
azimuthal wavenumber $m=1$. 

The rich variety of nonaxisymmetric instabilities can be demonstrated by
the addition of a differential rotation.  In this case even the
current-free (within the fluid) profile $B_\phi\propto 1/R$ (which
according to (\ref{tay}) is stable for $\Omega=0$) can become unstable.
Even for a differential rotation that by itself would be stable according
to \ref{ray}, the combination of $\Omega$ and $B_\phi\propto 1/R$ can be
unstable to $m=1$ perturbations (see Fig. \ref{AMRI}). We have called this
phenomenon the Azimuthal MagnetoRotational Instability (AMRI). It has even been  demonstrated that it should  be possible to observe
the AMRI in laboratory experiments, \cite{H09}.

Further new phenomena appear if an axial field is added, yielding a spiral,
or helical total field. In this case only a sufficient condition for
stability against axisymmetric perturbations is known. In the absence
of rotation this is 
\beg
 \frac{d}{dR}(R^2 B_\phi^2) < 0,
\label{mix0}
\ende
(see also Eq.~(\ref{chandra})). Including rotation, this was extended
\cite{HG62} to
\beg
R\frac{d\Omega^2}{dR} - \frac{1}{\mu_0 \rho R^3}
\frac{d}{dR}(RB_\phi)^2 > 0.
\label{mix}
\ende  
For the current-free field $B_\phi\propto 1/R$, only superrotating flows
with ${\rm d}\Omega/{\rm d} R>0$ are stable. Indeed, we have demonstrated
that dissipative Taylor-Couette flows beyond the Rayleigh limit for
centrifugal instability can easily be destabilized by helical  magnetic
fields with such a current-free azimuthal component \cite{HR05}. The
resulting axisymmetric traveling wave instability has become known as
the Helical MagnetoRotational Instability (HMRI), and has been obtained
in the PROMISE experiment \cite{R06,St06}.

In the PROMISE experiment the azimuthal field is $B_\phi\propto 1/R$.
In this paper we will also consider the generalization to $B_\phi =
a_B R + b_B/R$, where the extra term $a_B R$ corresponds to an axial
electric current running through the fluid as well, and hence opens the
possibility of current-induced (Tayler) instabilities. The resulting
(nonaxisymmetric) instabilities may also be modified by adding either
a differential rotation or an axial magnetic field.

One might suppose that adding an axial field would be important only if
its amplitude is of the same order as that of the azimuthal field.
Chandrasekhar \cite{C61} showed that for $\Omega=0$, a sufficiently strong
axial field will always suppress any axisymmetric instabilities of an
azimuthal field, by deriving the stability condition
\beg
I  B^2_z > \int \frac{\xi_R^2}{R^2} \frac{{\rm d}}{{\rm d}R}(R^2 B_\phi^2)
\ \ {\rm d}R,
\label{chandra}
\ende
where $I>0$ and $\xi_R$ is the (purely real) radial eigenfunctions. (Note
how (\ref{chandra}) reduces to (\ref{mix0}) for $B_z=0$.)  However, we
will show that the influence of $B_z$ cannot be ignored even for rather
small values.

We will find that, depending on the magnitudes of the imposed differential
rotation and magnetic fields, the field may either stabilize or destabilize the differential rotation, and the most unstable mode may be either the axisymmetric
Taylor vortex flow (the SMRI or HMRI), or the nonaxisymmetric AMRI, or the
nonaxisymmetric Tayler instability.  In combined axial and azimuthal fields,
we will also show that the nonaxisymmetric modes differ between $m$ and $-m$,
corresponding to left and right spirals.  As first pointed out by \cite{K96},
if the imposed field has both axial and azimuthal components, the system no
longer exhibits $\pm z$ symmetry.  For axisymmetric modes, the consequence of
this is that what were previously stationary modes (SMRI) become oscillatory,
traveling wave modes (HMRI).  For nonaxisymmetric modes, breaking the $\pm z$
symmetry of the basic state breaks the $\pm m$ symmetry of the instabilities.
Physically this corresponds to the fact that modes spiraling either in the
same or the opposite sense to the spiral structure of the basic state are
indeed different.  This $\pm m$ symmetry breaking is also a convenient
distinguishing feature between the AMRI and the Tayler instabilities; for the
AMRI the most unstable mode spirals in the opposite sense to the imposed field,
for the Tayler instabilities in the same sense.

Finally, in order to produce benchmarks for the application of incompressible
3D MHD codes, in this work we will focus primarily on magnetic Prandtl number
${\rm Pm}=1$.

%%%%%%%%%%%%%%%%%%%%%%%%%%%%%%%%%%%%%%%%%%%%%%%%%%%%%%%%%%%%%%%%%%%%%%%%%%%%%%%%%%%%%%%
\section{The equations}
%%%%%%%%%%%%%%%%%%%%%%%%%%%%%%%%%%%%%%%%%%%%%%%%%%%%%%%%%%%%%%%%%%%%%%%%%%%%%%%%%%%%%%%%
We are interested in the linear stability of the background field
$\vec{B}= (0, B_\phi(R), B_0)$, with $B_0=$const, and the flow
$\vec{U}= (0,R\Omega(R), 0)$.
The perturbed state of the system is  described by
\beg
u_R, \ u_\phi, \ u_z, \ p, \ b_R, \ b_\phi, \ b_z.
\ende
Developing the disturbances into normal modes, the solutions
of the linearized MHD equations are considered in the form
\beg
F=F(R){\textrm{exp}}({\textrm{i}}(kz+m\phi+\omega t)),
\label{nmode}
\ende
where $F$ is any of the velocity, pressure, or magnetic field disturbances.

The governing equations are
\begin{eqnarray}
\frac{\partial \vec{u}}{\partial t} + (\vec{U}\cdot\nabla)\vec{u}
 +  (\vec{u}\cdot\nabla)\vec{U}=
-\frac{1}{\rho} \nabla p + 
\nu \Delta \vec{u} + \\
+\frac{1}{\mu_0\rho}{\textrm{curl}}\ \vec{b} \times \vec{B}
+\frac{1}{\mu_0\rho}{\textrm{curl}}\ \vec{B} \times \vec{b},
\label{mhd}
\end{eqnarray}
\begin{eqnarray}
\frac{\partial \vec{b}}{\partial t}= {\textrm{curl}} (\vec{u} \times \vec{B})+  {\textrm{curl}} (\vec{U} \times \vec{b})+\eta \Delta\vec{b},
\label{mhd1}
\end{eqnarray}
and
\beg
{\textrm{div}}\ \vec{u} = {\textrm{div}}\ \vec{b} = 0,
\label{mhd2}
\ende
where $\vec{u}$ is the perturbed velocity, $\vec{b}$ the magnetic field,
$p$ the pressure perturbation.  $\nu$ is the kinematic viscosity and $\eta$
the magnetic diffusivity.
\begin{figure}[htb]
\psfig{figure=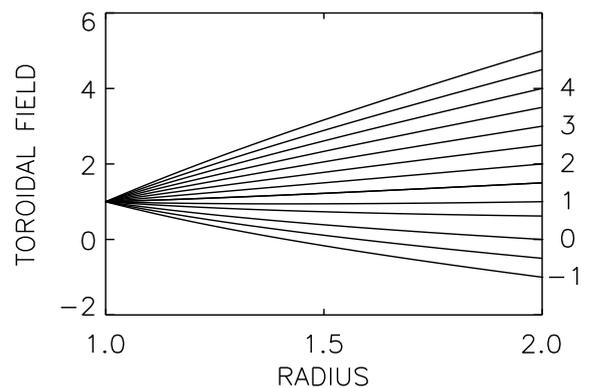,width=8.5cm,height=6cm}
\caption{\label{Bphi}
The basic state azimuthal field with prescribed values at the inner and
outer cylinders.
}
\end{figure}

The  stationary background solution is 
\beg
\Omega=a_\Omega +\frac{b_\Omega}{R^2},  \ \ \ \ \
B_\phi=a_B R+\frac{b_B}{R},
\label{basic}
\ende
where $a_\Omega$, $b_\Omega$, $a_B$ and $b_B$ are constants defined by 
\beg
a_\Omega=\Omega_{\rm{in}}\frac{ \mu_\Omega-{\hat\eta}^2}{1-{\hat\eta}^2}, \q
b_\Omega=\Omega_{\rm{in}} R_{\rm{in}}^2 \frac{1-\mu_\Omega}{1-{\hat\eta}^2},
\nonumber \\
a_B=\frac{B_{\rm{in}}}{R _{\rm{in}}}\frac{\hat \eta
( \mu_B - \hat \eta)}{1- \hat \eta^2},  \q
b_B=B_{\rm{in}}R _{\rm{in}}\frac{1-\mu_B \hat\eta}
{1-\hat \eta^2},
\label{ab}
\ende
with
\begin{equation}
\hat\eta=\frac{R_{\rm{in}}}{R_{\rm{out}}}, \; \; \;
\mu_\Omega=\frac{\Omega_{\rm{out}}}{\Omega_{\rm{in}}},  \; \; \;
\mu_B=\frac{B_{\rm{out}}}{B_{\rm{in}}}.
\label{mu}
\end{equation}
$R_{\rm{in}}$ and $R_{\rm{out}}$ are the radii of the inner and outer
cylinders, $\Omega_{\rm{in}}$ and $\Omega_{\rm{out}}$ are their rotation
rates, and $B_{\rm{in}}$ and $B_{\rm{out}}$ the azimuthal magnetic fields
at the inner and outer cylinders. The possible magnetic field solutions are
plotted in Fig. \ref{Bphi}.  Note that -- unlike $\Omega$, where
$\Omega_{\rm{in}}$ and $\Omega_{\rm{out}}$ are the physically relevant
quantities -- for $B_\phi$ the fundamental quantities are not so much
$B_{\rm{in}}$ and $B_{\rm{out}}$, but rather $a_B$ and $b_B$ themselves.
In particular, a field of the form $b_B/R$ is generated by running an
axial current only through the inner region $R<R_{\rm{in}}$, whereas a
field of the form $a_B R$ is generated by running a uniform axial current
through the entire region $R<R_{\rm{out}}$, including the fluid.  

Given the $z$-component of the electric current, ${\rm curl}_z \vec{B}= 2a_B$,
one finds for the current helicity of the background field
\begin{equation}
{\rm curl}\vec{B}\cdot \vec{B} =2 a_B B_0,
\label{ch}
\end{equation}
which may be either positive or negative (and of course vanishes for the
current-free case $a_B=0$).  However, both signs yield the same instability
curves, merely with the previously mentioned left and right spirals
interchanged.

The inner value $B_{\rm in}$ is normalized with the uniform vertical field,
i.e.
\begin{equation}
\beta =\frac{B_{\rm in}}{B_0}.
\label{beta}
\end{equation}
For $\mu_B=1$ we have
\begin{equation}
{\rm curl}\vec{B}\cdot \vec{B} =\frac{2\beta}{3}   \frac{B_0^2}{R_{\rm in}}.
\label{bet}
\end{equation}
The sign of $\beta$ thus determines the sign of the helicity of the background
field. And again, interchanging $\pm\beta$ simply interchanges left and right
spirals $\pm m$.

As usual, the toroidal field amplitude is measured by the  Hartmann number
\beg
{\rm Ha} = \frac{B_{\rm in} R_0}{\sqrt{\mu_0 \rho \nu \eta}}.
\label{Ha}
\ende
$R_0=\sqrt{R_{\rm in}(R_{\rm out} - R_{\rm in})}$ is used as the unit of length,
$\eta/R_0$ as the unit of velocity and $B_{\rm in}$ as the unit of the azimuthal
fields. Frequencies, including the rotation $\Om$, are normalized with the inner rotation rate $\Om_{\rm in}$. The magnetic Reynolds number Rm is defined as 
\beg
{\rm Rm}=\frac{\Om_{\rm in}  R_0^2}{\eta}.
\label{Rey}
\ende
The Lundquist number S is defined by $\rm S=Ha \cdot \sqrt{Pm}$.

The boundary conditions associated with the perturbation equations are no-slip
for $\vec{u}$,
\beg
u_R=u_\phi=u_z=0,
\label{ubnd}
\ende
and perfectly conducting for $\vec{b}$,
\beg
db_\phi/dR + b_\phi/R = b_R = 0.
\label{bcond}
\ende
These boundary conditions hold for both $R=R_{\rm{in}}$ and $R=R_{\rm{out}}$.

If we consider a constant phase of a nonaxisymmetric pattern,
Eq.~(\ref{nmode}) yields
\beg
\frac{\partial z}{\partial t}\bigg\vert_{\phi}= - \frac{\Re(\omega)}{k}, \ \ \ \ \ \ \ \ \  \frac{\partial \phi}{\partial t}\bigg\vert_{z}= - \frac{\Re(\omega)}{m}.
\label{re}
\ende
The first relation describes the phase velocity of the modes in the axial
direction, the second in the azimuthal direction (and only exists for
nonaxisymmetric modes). Obviously the wave is traveling upwards if the real
part of the eigenfrequency is negative.

At a fixed time the phase relations (\ref{re}) can also be written as 
\beg
\partial z/\partial \phi = - m/k. 
\label{hel}
\ende
Now, $k$ and $m$ are both real numbers, and without loss of generality one
of them can be taken to be positive, say $k$.  The other one, $m$ in this
case, must be allowed to have both signs though. Negative $m$ describe
right-hand spirals (marked here by R), and positive $m$ describe left-hand
spirals (marked here by L). If the axisymmetric background field possesses
positive $B_z$ and $B_\phi$ (as used for the calculations here) then its
current helicity is positive, or equivalently, it forms a right spiral.
%%%%%%%%%%%%%%%%%%%%%%%%%%%%%%%%%%%%%%%%%%%%%%%%%%%%%%%%%%%%%%%%%%%%%%%%%%
\section{From AMRI to HMRI}
\begin{figure}[htb]
\psfig{figure=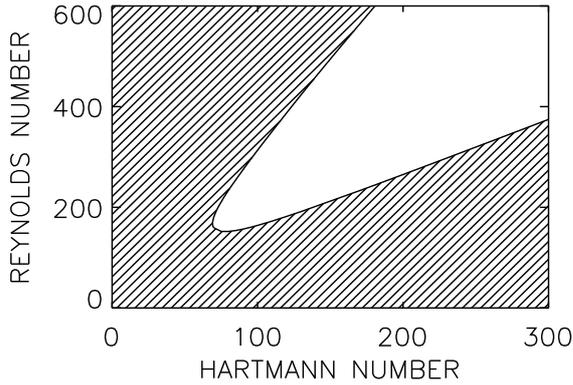,width=8.5cm,height=6cm}
\caption{\label{AMRI}
Azimuthal  MagnetoRotational Instability (AMRI):
The instability curve for current-free azimuthal magnetic fields.
The most unstable mode is {\em nonaxisymmetric}, with $m=1$. Axisymmetric
instabilities do not exist at all in this configuration.
 $\mu_B=0.5$, $\mu_\Omega=0.5$. $\rm Pm=1$.}
\end{figure}

%%%%%%%%%%%%%%%%%%%%%%%%%%%%%%%%%%%%%%%%%%%%%%%%%%%%%%%%%%%%%%%%%%%%%%%%
We begin with a purely azimuthal field, and no electric currents within
the fluid, that is, $B_\phi \propto 1/R$.  Figure \ref{AMRI} presents results
for $\mu_\Omega=0.5$, showing that for $\rm Ha>100$ and $\rm Re>200$ there
exists an $m=1$ nonaxisymmetric instability.  Note also how both the upper
and lower branches of the instability curve tilt to the right, that is,
have a positive slope $ d{\rm Re}/d{\rm Ha}$. For a given Hartmann number,
the instability therefore only exists within a finite range of Reynolds
numbers.  If $\rm Re$ is too large, the instability disappears again as a
consequence of the suppressing action that differential rotation often has
on nonaxisymmetric modes.

\begin{figure}[htb]
\psfig{figure=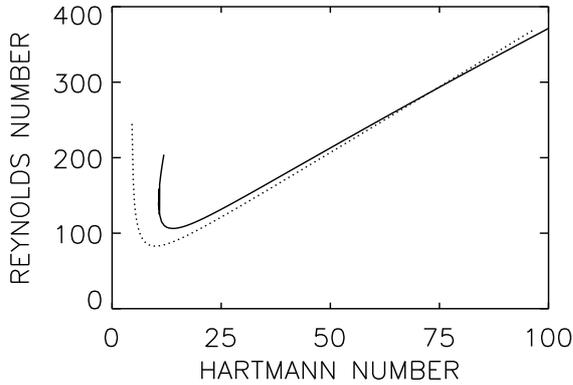,width=8.5cm,height=6cm}
\caption{\label{SMRI}
Standard MRI: The instability curves for uniform axial magnetic fields.
For ${\rm Ha}<75$ the axisymmetric mode is preferred; for ${\rm Ha}>75$ the
$m=1$ nonaxisymmetric mode is preferred. The global minimum $\rm Re$ value
is for the axisymmetric mode. $\mu_\Omega=0.5$, $\rm Pm=1$.}
\end{figure}

We consider next a purely axial field, the so-called standard MRI.  In this
case both axisymmetric and nonaxisymmetric modes may be excited, with the
axisymmetric mode being the one with the overall lowest Reynolds number
(Fig. \ref{SMRI}). For $\rm Pm=1$ this overall minimum occurs for
$\rm Ha\simeq 10$ and $\rm Re\simeq 80$.  However, for sufficiently large
$\rm Ha$ the $m=1$ nonaxisymmetric mode is actually preferred over the
axisymmetric mode. The standard MRI for purely axial fields is therefore
not necessarily an axisymmetric mode. The axisymmetric mode only dominates
for sufficiently weak fields, including also the global minimum $\rm Re$
value. It also dominates the entire weak-field branch of the instability 
curves (Fig.~\ref{SMRI}). The axisymmetric mode here tilts to the left,
whereas the nonaxisymmetric mode tilts to the right, as before in Fig.
\ref{AMRI}.

\begin{figure}[htb]
\vbox{
\psfig{figure=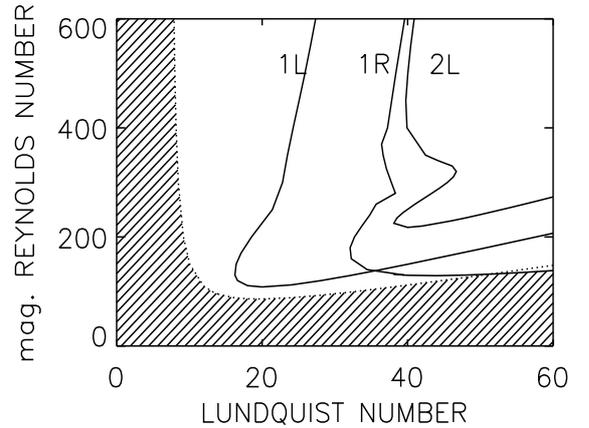,width=8.5cm,height=6.5cm}
\psfig{figure=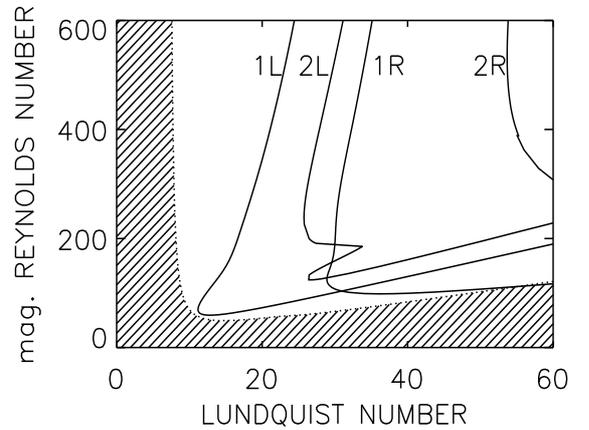,width=8.5cm,height=6.5cm}}
\caption{\label{HMRI} Helical magnetorotational instability (HMRI): The
instability curves for current-free ($\mu_B=0.5$) helical fields with
$\beta=2$. $\mu_\Omega=0.5$.  The dotted line gives the axisymmetric mode.
The solid lines are marked with the mode number $m$ and the type of helicity
(L left, R right). Note that for stronger fields the most unstable mode is
nonaxisymmetric with the same helicity as the helicity of the background
field. The mode with the lowest Re is always {\em axisymmetric}.
Top: $\rm Pm=1$, bottom: $\rm Pm=0.01$.}
\end{figure}

Figure \ref{HMRI} finally shows results combining azimuthal and axial fields,
focusing in this case on $\beta=2$ (so a right-handed helicity).  We see the
same general pattern as before: only the weak-field branch of the $m=0$ mode
tilts to the left; all nonaxisymmetric modes tilt to the right.  Up to
${\rm Ha}\approx50$ the axisymmetric mode is preferred, just as before for
the standard MRI. For ${\rm Ha}>50$ the $m=1$ right spiral is preferred.

Figure \ref{HMRI} also demonstrates that the (axisymmetric) standard MRI and
the (nonaxisymmetric) AMRI are the basic elements which both appear, with
different weights, if the background field has a spiral geometry. From this
point of view instabilities in helical fields are simply a mixture of these
two basic elements.  More specifically, one finds that the weak-field branch
of the instability in Fig.~\ref{HMRI} is very similar to the weak-field
branch of the standard MRI (Fig.~\ref{SMRI}) while the strong-field branch
strongly resembles the strong-field branch of AMRI (Fig.~\ref{AMRI}). The
minimum is always obtained for the axisymmetric mode of the standard MRI.
The only difference between
the standard MRI and HMRI is the different character of the eigenfrequencies:
the SMRI is stationary, whereas the HMRI is oscillatory, as a necessary
consequence of the $\pm z$ symmetry-breaking \cite{K96}. It is precisely
this oscillatory nature of the HMRI that has been used to identify it in
the PROMISE experiment \cite{R06,St06}.

Note finally that taking $\rm Pm=1$ greatly simplifies the results, and
indeed eliminates some particularly interesting results.  As we have
previously demonstrated, both the (axisymmetric) HMRI \cite{HR05} as well
as the (nonaxisymmetric) AMRI \cite{H09} have the property that their
scalings with Pm vary dramatically with $\mu_\Omega$. For $\mu_\Omega$
only somewhat greater than the Rayleigh value, both modes have Ha and Re as
the relevant measures of field strength and rotation rates, whereas for
greater values of $\mu_\Omega$, $\rm S=Ha\cdot\sqrt{Pm}$ and $\rm Rm=
Re\cdot Pm$ are the relevant measures.  For small Pm the differences can
thus be huge. Insulating versus conducting boundaries can also have a
surprisingly large influence on this transition from one scaling to another
\cite{RH07}.
%%%%%%%%%%%%%%%%%%%%%%%%%%%%%%%%%%%%%%%%%%%%%%%%%%%%%%%%%%%%%%%%%%%%%%%%%%
\section{Fields with current-helicity}
\subsection{Steep rotation law}
%%%%%%%%%%%%%%%%%%%%%%%%%%%%%%%%%%%%%%%%%%%%%%%%%%%%%%%%%%%%%%%%%%%%%%%%
\begin{figure}[htb]
\psfig{figure=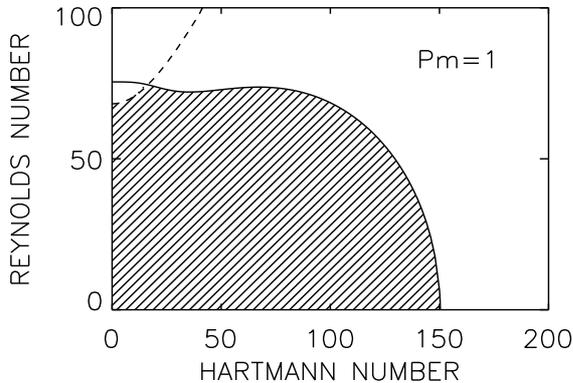,width=8.5cm,height=6cm}
\caption{\label{TI} Tayler instability (TI) of a toroidal field under the
influence of differential rotation with stationary outer cylinder.
The solid curve is $m=1$, the dotted curve is $m=0$.
$\mu_B=1$, $\mu_\Omega=0$. $\rm Pm=1$.}
\end{figure}

We begin by considering the stability of purely toroidal fields, and
differential rotation profiles with a stationary outer cylinder. There
are then three classical results known: First, in the absence of any
fields, axisymmetric Taylor vortices arise at $\rm Re=68$, and nonaxisymmetric
instabilities at $\rm Re=75$. Second, in the absence of any rotation, $m=1$
Tayler instabilities arise at $\rm Ha=150$. Figure \ref{TI} shows how these
results are linked when both Ha and Re are non-zero. For Ha very small, the
axisymmetric Taylor vortex mode is stabilized, whereas the nonaxisymmetric
mode is eventually destabilized, and connects smoothly to the pure Tayler
instability.

\begin{figure}[h]
\vbox{
\psfig{figure=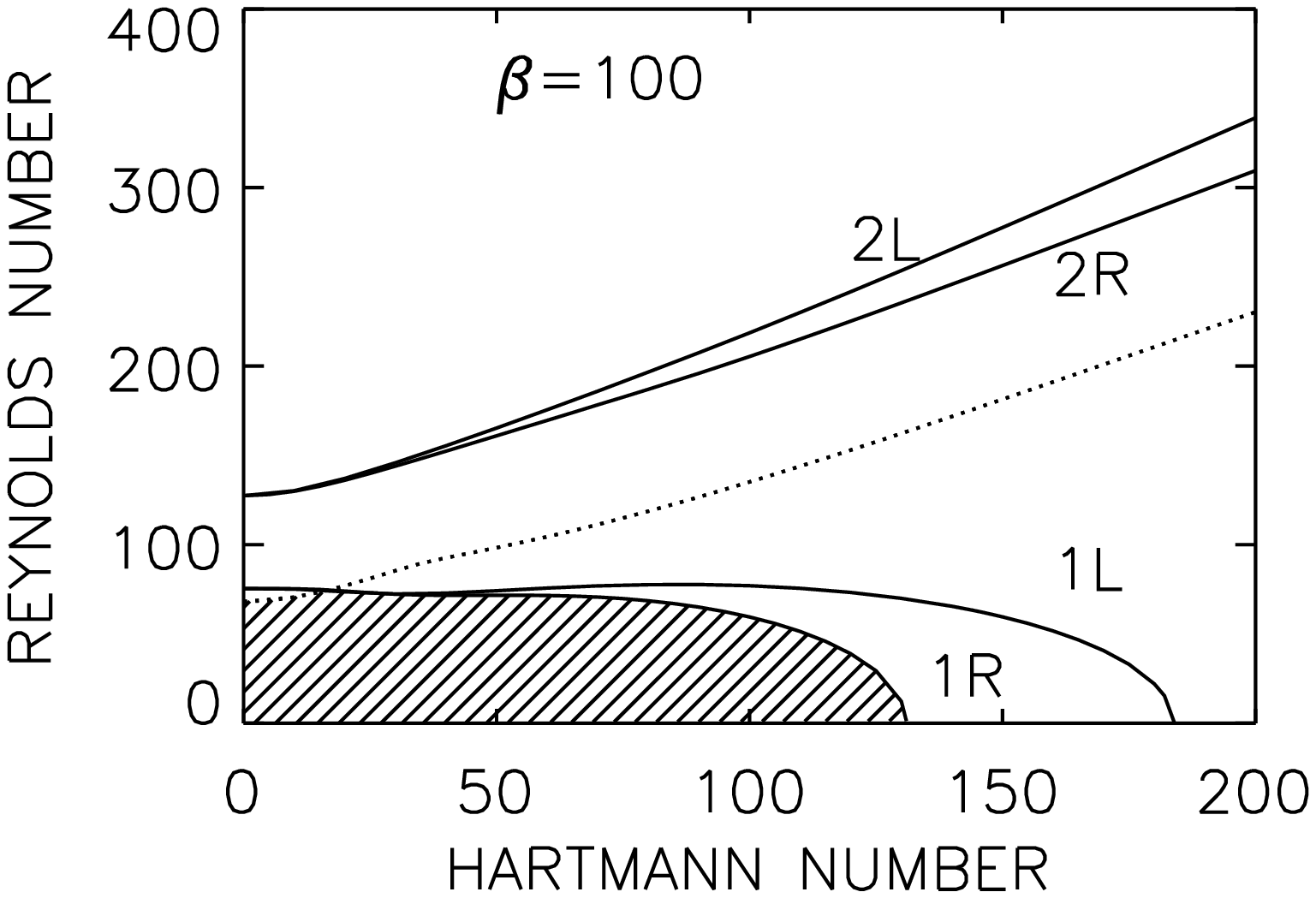,width=8.5cm,height=5.5cm}
\psfig{figure=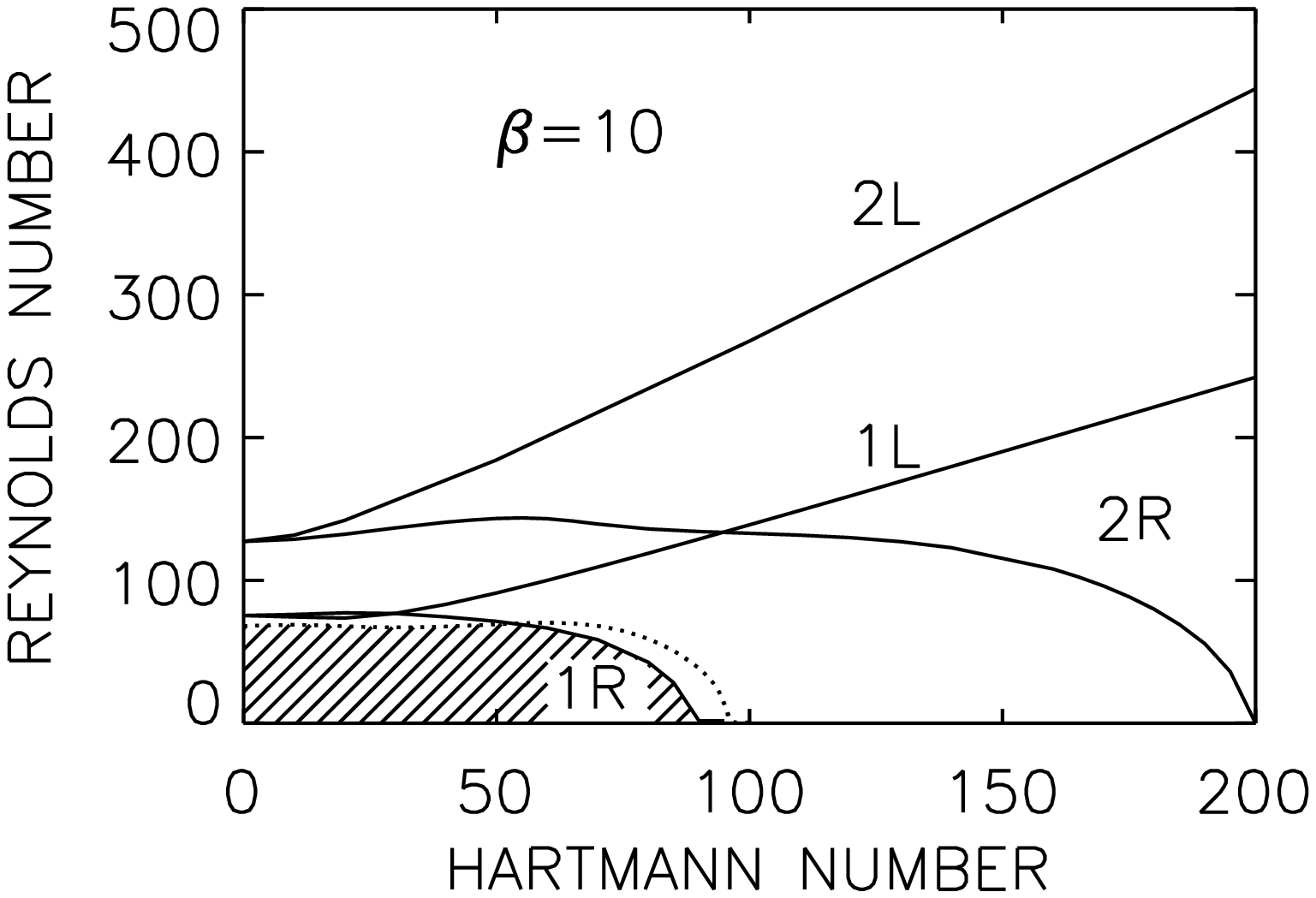,width=8.5cm,height=5.5cm}
\psfig{figure=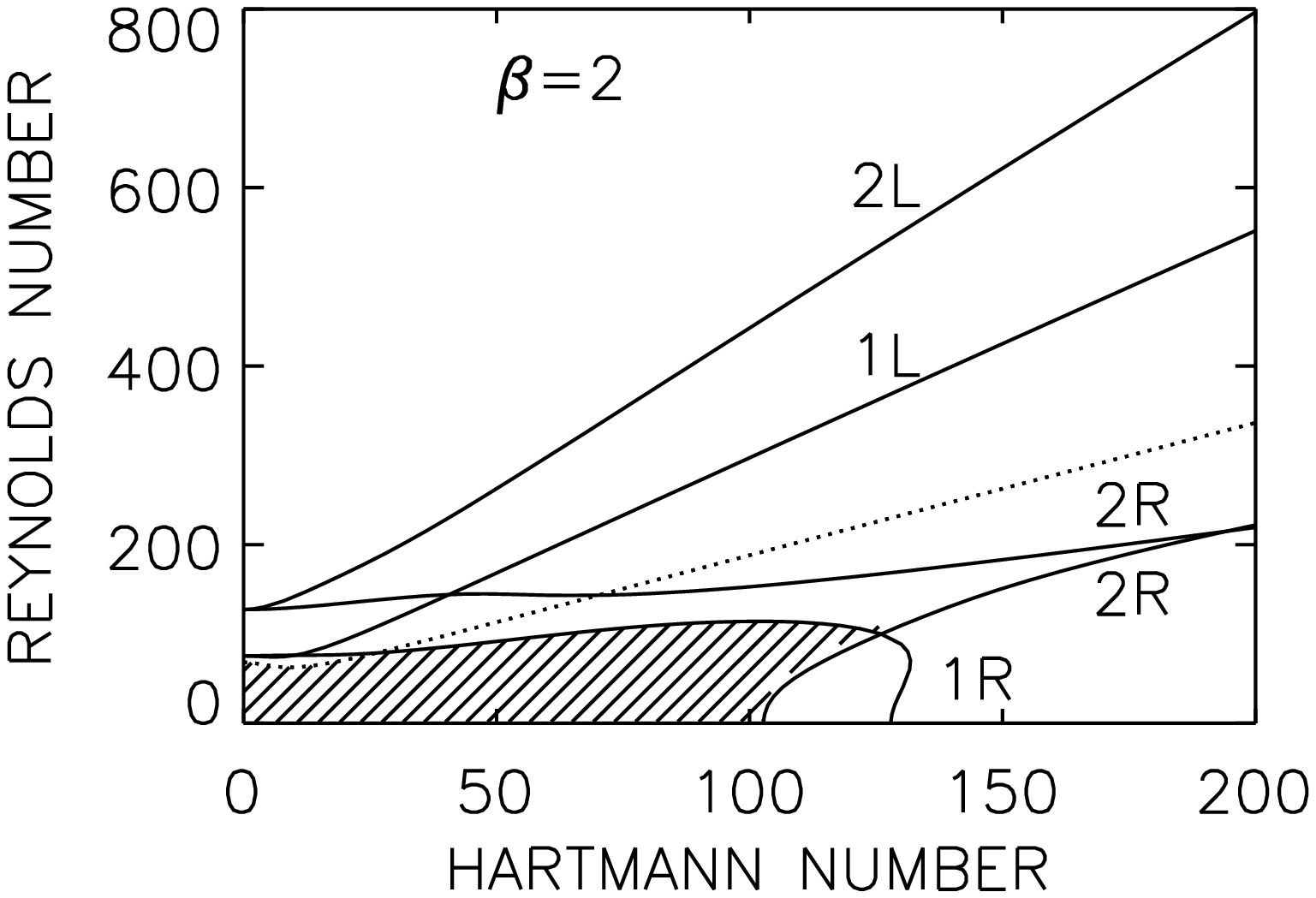,width=8.5cm,height=5.5cm}
}
\caption{\label{f1} Instability curves for magnetic fields with positive current
helicity. The curves are marked with their azimuthal wave numbers $m$. The shaded
areas are the stable regions. Without magnetic fields the curves always start
at $\rm Re=68$ for the $m=0$ Taylor vortices. Note the dominance of the
axisymmetric modes (dotted) also for weak magnetic fields, unless for very high $\beta$ the
field becomes nearly toroidal. $\rm Pm=1$, $\mu_B=1$, $\mu_\Omega=0$.}
\end{figure}

The next step is to add a uniform axial field to the azimuthal field, with
(say) positive polarity. The background field then has a positive helicity,
that is, it spirals to the right. If the axial field is weak, e.g. with
$\beta=100$, then the marginal instability curves (Fig. \ref{f1}, top) strongly
resemble the map for $\beta\to\infty$ (Fig. \ref{TI}). The main differences are
i) the slightly smaller Hartmann number of the toroidal field, and ii) the
splitting of the spiral modes $m=1$ and $m=-1$ into two curves with different
helicity (R and L). The left-hand modes require a greater rotation than the
right-hand modes. For background fields with positive helicity, we thus find
that the right spirals are preferred, whereas for background fields with negative
helicity, R and L would be exchanged, and the left spirals would be preferred.

For $\beta=10$ the differences between the L and R modes for given $m$ increase,
so that the pure Tayler instability exists only as the 1R mode. The 1L mode no
longer connects to $\rm Re=0$, and is not the most unstable mode anywhere in
the given domain (Fig. \ref{f1}, middle). 

The 1R mode also dominates for $\beta$ of order unity. There is, however, an interesting particularity in this case. For very slow rotation, a 2R mode
reduces the stability domain. For $\rm Re\simeq 0$, and in a limited range of
Ha ($\rm Ha\simeq 100...130$), this mode forms the first instability
(see \cite{BU08}). A small amount of differential rotation, however, brings
the system back to the 1R instability.

For models with helical fields and steep rotation laws (with stationary
outer cylinder), we indeed find the expected splitting between right and left
spiral instabilities. If the axisymmetric background field is right-handed,
then the first unstable mode is also right-handed. The corresponding critical
magnetic field strength is reduced compared to the TI of purely toroidal
fields. If both magnetic field components are of the same order then the 2R
mode is found to destabilize the system at the strong-field side of the
stability domain, but only for very slow rotation. The differential rotation basically limits the action of this 2R mode.

%%%%%%%%%%%%%%%%%%%%%%%%%%%%%%%%%%%%%%%%%%%%%%%%%%%%%%%%%%%%%%%%%%%%%
\subsection{Flat rotation law}
%%%%%%%%%%%%%%%%%%%%%%%%%%%%%%%%%%%%%%%%%%%%%%%%%%%%%%%%%%%%%%%%%%%%%%%

\begin{figure}[h]
\psfig{figure=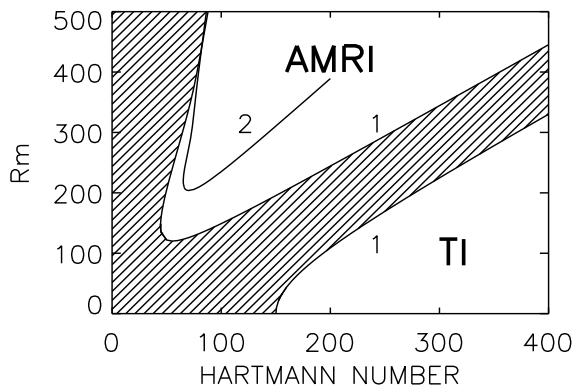,width=8.5cm,height=6cm}
\caption{\label{AMRITI} Flat rotation law ($\mu_\Omega=0.5$) and  $B_z=0$. The curves are marked with their mode number $m$. The AMRI occurs in the upper left
corner, and the TI in the lower right corner. $\mu_B=1$, $\rm Pm=1$}
\end{figure}

For weak magnetic fields and the steep rotation law, the axisymmetric Taylor
vortex mode is the most easily excited instability. For a sufficiently flat
rotation law the non-magnetic Taylor vortices necessarily disappear, and a
critical Reynolds number no longer exists for $\rm Ha=0$. For the flat
rotation law with $\mu_\Omega=0.5$, and the nearly uniform toroidal field
with $\mu_B=1$, the instability curves for purely toroidal fields are given
by Fig. \ref{AMRITI}. Both of the previous instabilities appear in this case:
TI exists in the lower right corner, and AMRI exists in the upper left corner.
The AMRI arises from the term $b_B/R$ in the magnetic background field profile
(\ref{basic}), while the current-driven TI is due to the term $a_B R$. The two
instabilities are separated by a stable branch with ${\rm Re}\approx {\rm Ha}$,
where the differential rotation stabilizes the TI. 
\begin{figure*}[t]
\mbox{
\vbox{
\psfig{figure=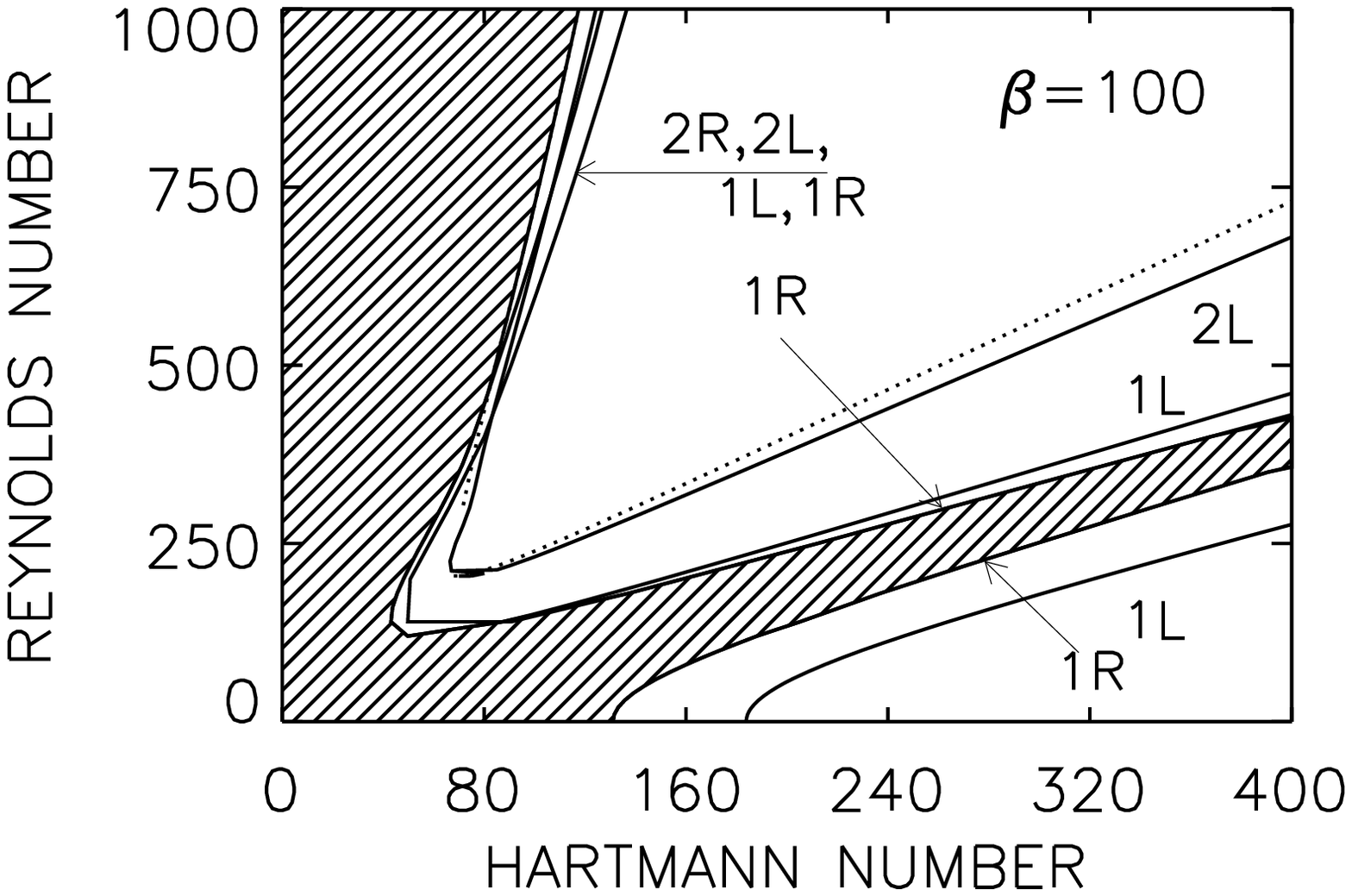,width=9cm,height=6cm}
\psfig{figure=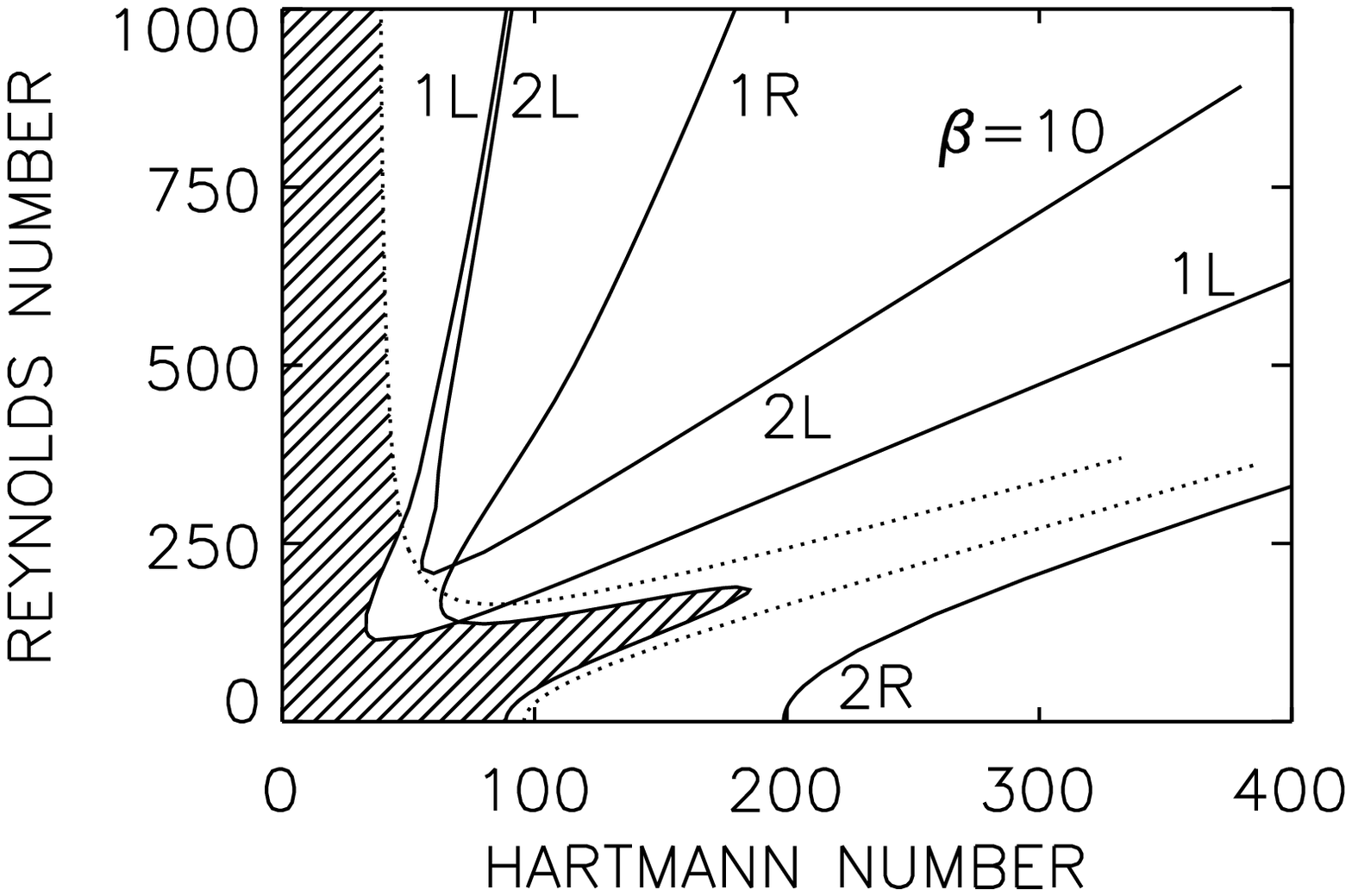,width=9cm,height=6cm}}
\vbox{
\psfig{figure=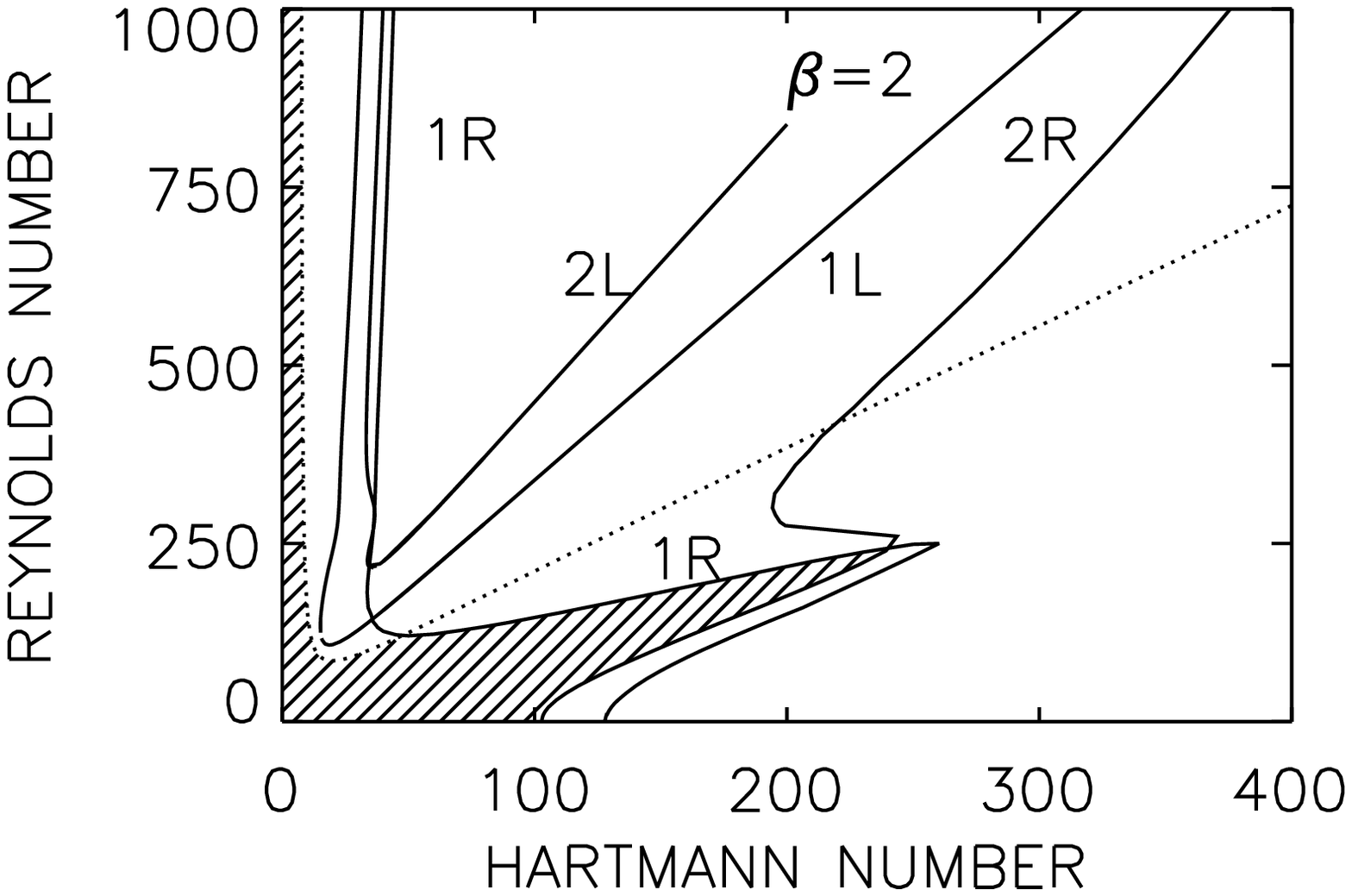,width=9cm,height=6cm}
\psfig{figure=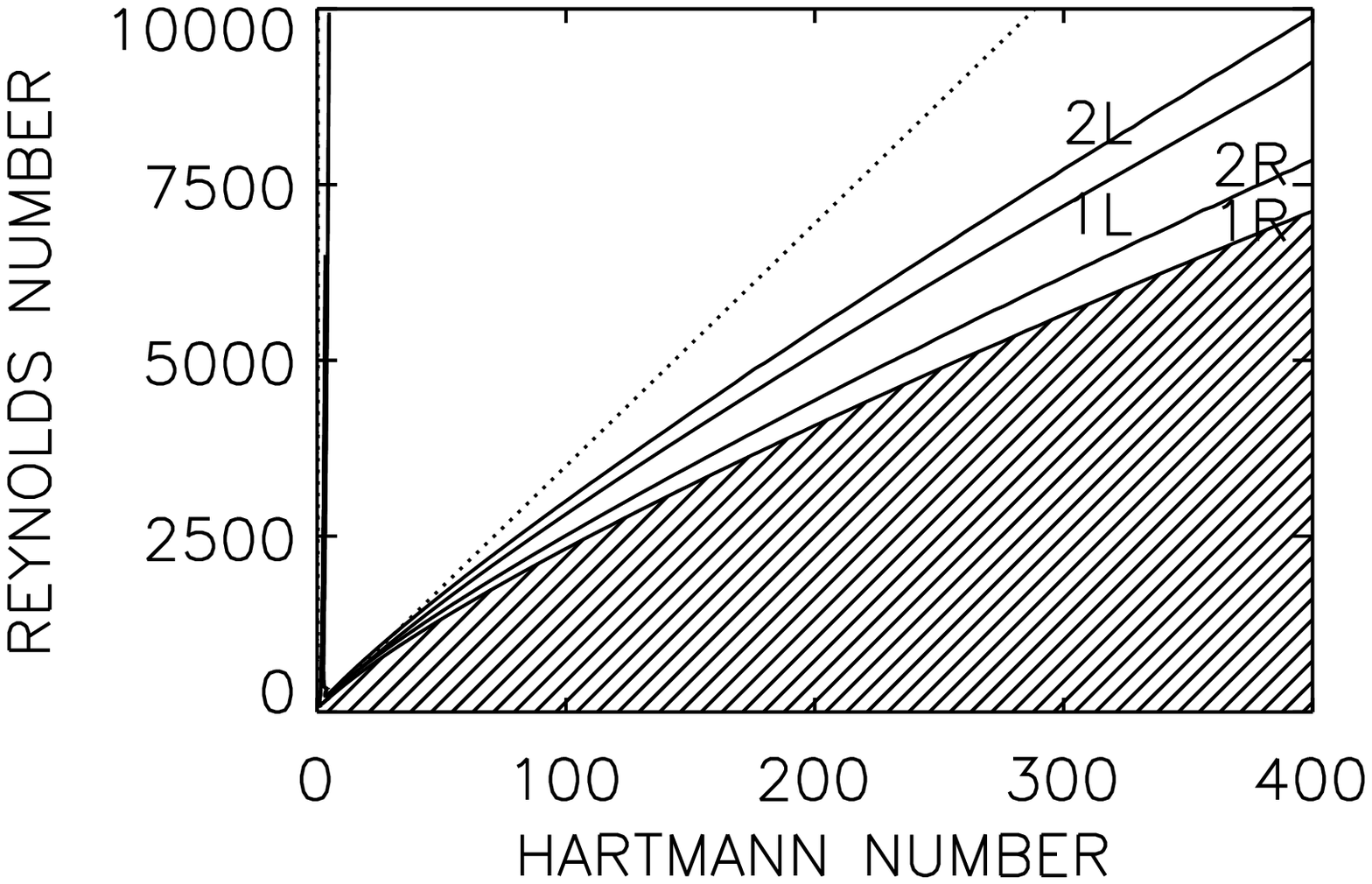,width=9cm,height=6cm}}
}
\caption{\label{f4} The same as in Fig.~\ref{AMRITI} but for finite values of $\beta$: Left: $\beta=100$ (top) and  $\beta=10$ (bottom), Right:  $\beta=2$ (top) and $\beta=0.1$ (bottom). The curves are marked with the azimuthal mode numbers $m$, the curves for $m=0$ are dotted. The notation R (right spiral) stands for  negative $m$ and the notation L (left spiral) stands for  positive $m$, It is $\mu_B=1$, $\rm Pm=1$.}
\end{figure*}

For $\beta=\infty$ the modes with positive and negative $m$ are degenerate.
At the weak-field limit the line for $m=2$ even crosses the line for $m=1$.
Nevertheless, the AMRI solution with the lowest Reynolds number is a {\em
nonaxisymmetric} mode with $m=1$. We find that this remains true for helical
background fields with large $\beta$, but for $\beta$ of order unity and smaller
 the $m=0$ mode yields the instability with the lowest Reynolds 
number (Fig. \ref{f4}) -- as is also true for the standard MRI and HMRI. The 
transition from nonaxisymmetry to axisymmetry can be accomplished simply by 
increasing the axial component of the background field. It is thus clear 
that there is a smooth transition from one form of the MRI in TC flows to the
next. The same is true for the corresponding eigenfrequencies, which develop
from real values (for standard MRI) to complex values (in all  other cases).

\begin{figure}[h]
\psfig{figure=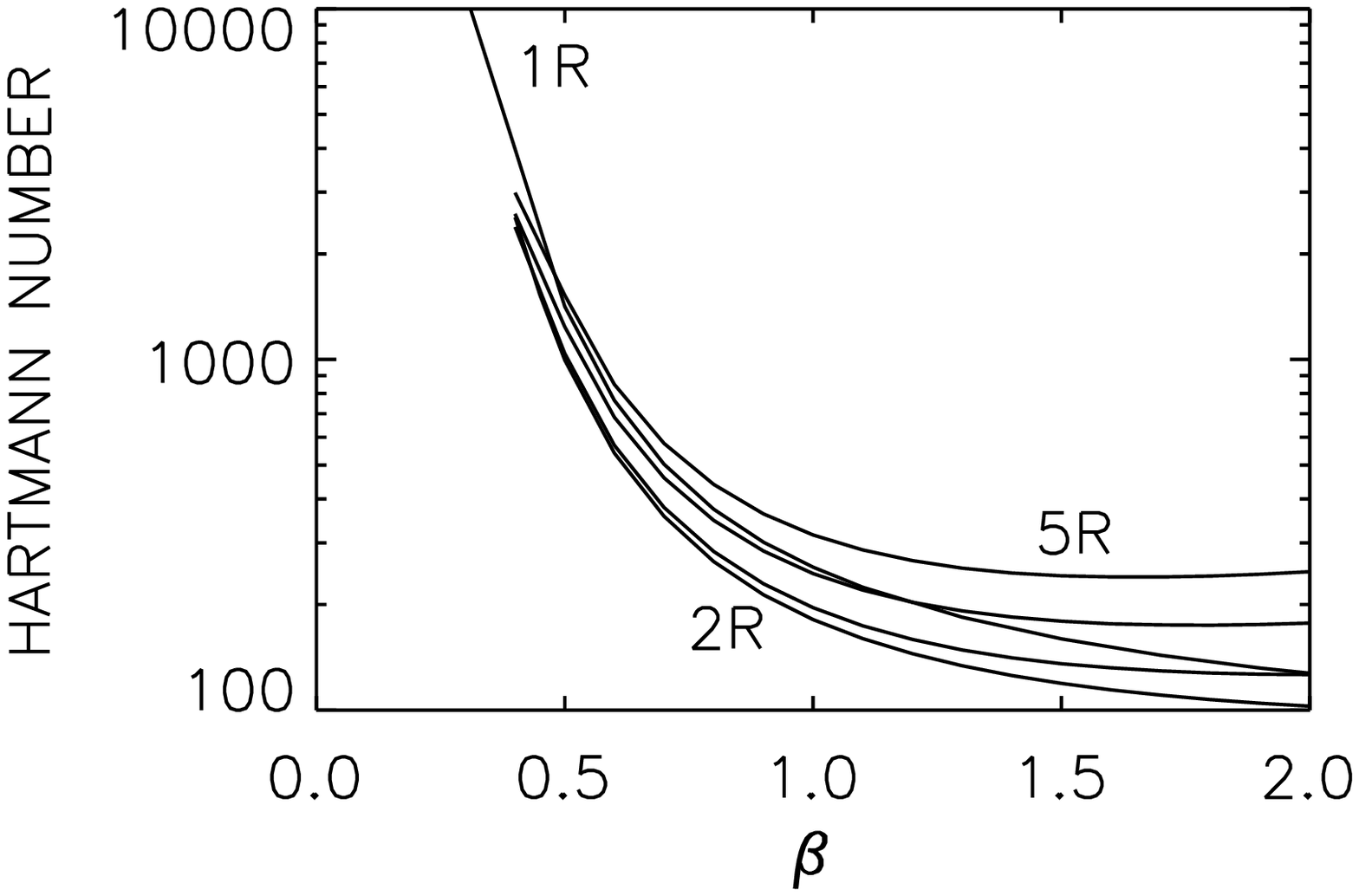,width=9cm,height=7cm}
\caption{\label{f4a} The critical Hartmann number (\ref{Ha}) for  $m =-1\dots -5$ and  $\beta \lsim 10$. An increasing dominance  of the axial magnetic field component acts {\em stabilizing}.  The curves do not depend on  $\rm Pm$. $\mu_B=1$.}
\end{figure}

On the other hand, if large-scale electric currents flow through the fluid,
a critical Hartmann number exists for $\rm Re=0$, similar to Fig. \ref{TI},
where the system is also unstable even for $\rm Re=0$. In this case the
critical Hartmann number is unchanged; it is again $\rm Ha=150$ for purely
toroidal fields, i.e. $\beta=\infty$ (Figs. \ref{TI}, \ref{AMRITI}). This
value does not even depend on the magnetic Prandtl number. For
increasing $\beta$, however, the critical Hartmann number is  reduced to
about 100. The most unstable mode is 1R for $\beta \geq 10$, but is again
2R for $\beta$ of order unity. This result holds for very weak differential
rotation; only then  a mode higher than $m=1$ plays a
role in the transition from stability to instability. 

For background fields
with positive helicity the TI favors instability patterns with right spirals.

The instability curves of the weak-field, or diffusion-dominated (AMRI) limit
also show a characteristic behavior. For large $\beta$ it is formed by the 
nonaxisymmetric modes, while for small $\beta$ the axisymmetric mode prevails.
Consequently, the slopes of the lines change from positive for the
nonaxisymmetric modes to negative for the axisymmetric modes (Fig. \ref{f4}). 
Again, the transition from AMRI to standard MRI becomes
clear by variation of $\beta$. If the preferred modes are nonaxisymmetric
(for large $\beta$), then the spirals are always left-handed. The different
mode pattern is the characteristic difference to the preferred modes in the
TI domain.
%\begin{figure*}[htb]
%\mbox{
%\psfig{figure=2m_beta2.eps,width=8.5cm,height=5.5cm}
%\psfig{figure=2m_beta-2.eps,width=8.5cm,height=5.5cm}
%}
%\caption{\label{f11} Radial component of the resulting magnetic pattern for a nonrotating container with $\rm Ha=200$. %LEFT: $\beta=2$, right spirals with $m=2$. RIGHT: $\beta=-2$, left spirals with $m=2$.
%$\mu_B=1$, $\rm Pm=1$. }
%\end{figure*}

%%%%%%%%%%%%%%%%%%%%%%%%%%%%%%%%%%%%%%%%%%%%%%%%%%%%%%%%%%%%%%%%%%%%%
\subsection{No rotation}
%%%%%%%%%%%%%%%%%%%%%%%%%%%%%%%%%%%%%%%%%%%%%%%%%%%%%%%%%%%%%%%%%%%%%%%
In general, for given Hartmann number the differential rotation stabilizes the Tayler instability which also exists without any rotation. On the other hand, 
we have shown that  the critical Hartmann numbers  for nonrotating containers  do not depend on the given value of the magnetic Prandtl number $\rm Pm$, \cite{RS10}. Hence, the results given in Fig. \ref{f4a} for $\rm Re=0$ and for  $m =-1\dots -5$ are also valid for the small magnetic Prandtl numbers of liquid metals such as sodium or gallium which are used in the laboratory.

The question about the critical Hartmann numbers for $\beta \lsim 1$ arises if the azimuthal 
mode number $m$ is varied.  Generally  the mode with $m=-2$ dominates but for $\beta \lsim 0.4 $ the mode 
with $m=-3$ starts to be preferred. It may happen that even higher $m$ appear to be preferred for 
even smaller $\beta$. However, it will  only happen for so high values of  the  Hartmann number ($\gsim 2500$) 
that  i) laboratory experiments are impossible and ii)  numerical investigations with differential  rotation included -- which in particular stabilizes higher $m$ -- are not possible.  The basic  
result of the calculations is that the reduction of the increase  of the axial field component  ($\beta \lsim 1$) 
acts strongly stabilizing. This the more as the normalized differences of the critical Hartmann numbers for various $m$ 
become smaller and smaller. These  results do not change if  formulated with the Hartmann 
number of the axial field rather than with  the  Hartmann number of the toroidal  field. The total energy which is necessary to excite TI strongly grows with decreasing $\beta$. Absolutely no  instability remains  for the limit $\beta \to 0$.  

We  know from previous calculations that for $\Omega=0$ an almost homogenous  toroidal field ($\mu_B=1$) becomes unstable against disturbances with azimuthal number $m=-1$ for $\rm Ha \geq 150$.  If an axial field is added then the critical Hartmann number is reduced, i.e. the toroidal field is destabilized by the axial component.  While for $B_z=0$ no preferred helicity exists for the instability pattern with axial field the resulting  spiral geometry is the same as that of the background field. We also find that a total minimum of the critical Hartmann number 
exists for $\beta\lsim 10$  (typical values of the experiment PROMISE) where the mode with $m=1$ is the most unstable one. If the axial field starts to dominate 
for   $\beta < 2$ then the critical Hartmann numbers are growing, i.e. system becomes more and more stable (see Fig. \ref{f4a}). 
%%%%%%%%%%%%%%%%%%%%%%%%%%%%%%%%%%%%%%%%%%%%%%%%%%%%%%%%%%%%%%%%%%%%%%%%%%%%%%%
\section{Nonlinear simulations}
%%%%%%%%%%%%%%%%%%%%%%%%%%%%%%%%%%%%%%%%%%%%%%%%%%%%%%%%%%%%%%%%%%%%%%%%%%%%%%%
The previous results have all been purely linear onset calculations, in
which the governing equations are reduced to a linear, one-dimensional
eigenvalue problem. It is also of interest to study the nonlinear
equilibration of some of these modes, which we do with a three-dimensional
spectral MHD code \cite{Ge07}.  The code is based on Fourier modes in
$\phi$; for each Fourier mode the $(R,z)$ structure is discretized by
standard spectral element methods involving Legendre polynomials \cite{Fournier}.
For Reynolds numbers only slightly beyond the linear onset, the solutions do not
develop much structure yet, so 16 Fourier modes were sufficient in $\phi$.
For the $(R,z)$ structure, we typically used 2 spectral elements in $R$,
and 14 in $z$ (where periodicity with a domain height of $\Gamma = 2\pi
(R_\mathrm{out} - R_\mathrm{in})$ was enforced), with a polynomial order
between 12 and 18.  The time-stepping uses third-order Adams-Bashforth for
the nonlinear terms, and second-order Crank-Nicolson for the diffusive terms.
Boundary conditions are as before, no-slip for $\vec U$, and perfectly
conducting for $\vec B$. Initial conditions are the basic Couette profile
for $\vec U$, and random perturbations of size $10^{-6}B_{\rm in}$ for
$\vec B$. 

We begin by verifying that transforming $\beta\to -\beta$ has the expected
result. 
%Figure \ref{f11} shows results for $\beta=\pm2$. 
Positive/negative
$\beta$ do indeed yield right/left mirror image spirals, verifying the linear
onset conclusion that these instabilities spiral in the same sense as the
imposed field.

The simulations concern the linear onset curves for flat rotation law (see
Fig. \ref{f4}) both for AMRI and TI. Two examples are given for each
instability, to probe the spatial pattern and the resulting field strength
of the modes. The helical structure of all solutions is clearly visible,
dominated by low Fourier modes $m=1$ and/or $m=2$ in agreement with the
linear analysis. The solutions are stationary, except for a drift in the
azimuthal direction.

\begin{figure*}[htb]
\mbox{
\vbox{
\psfig{figure=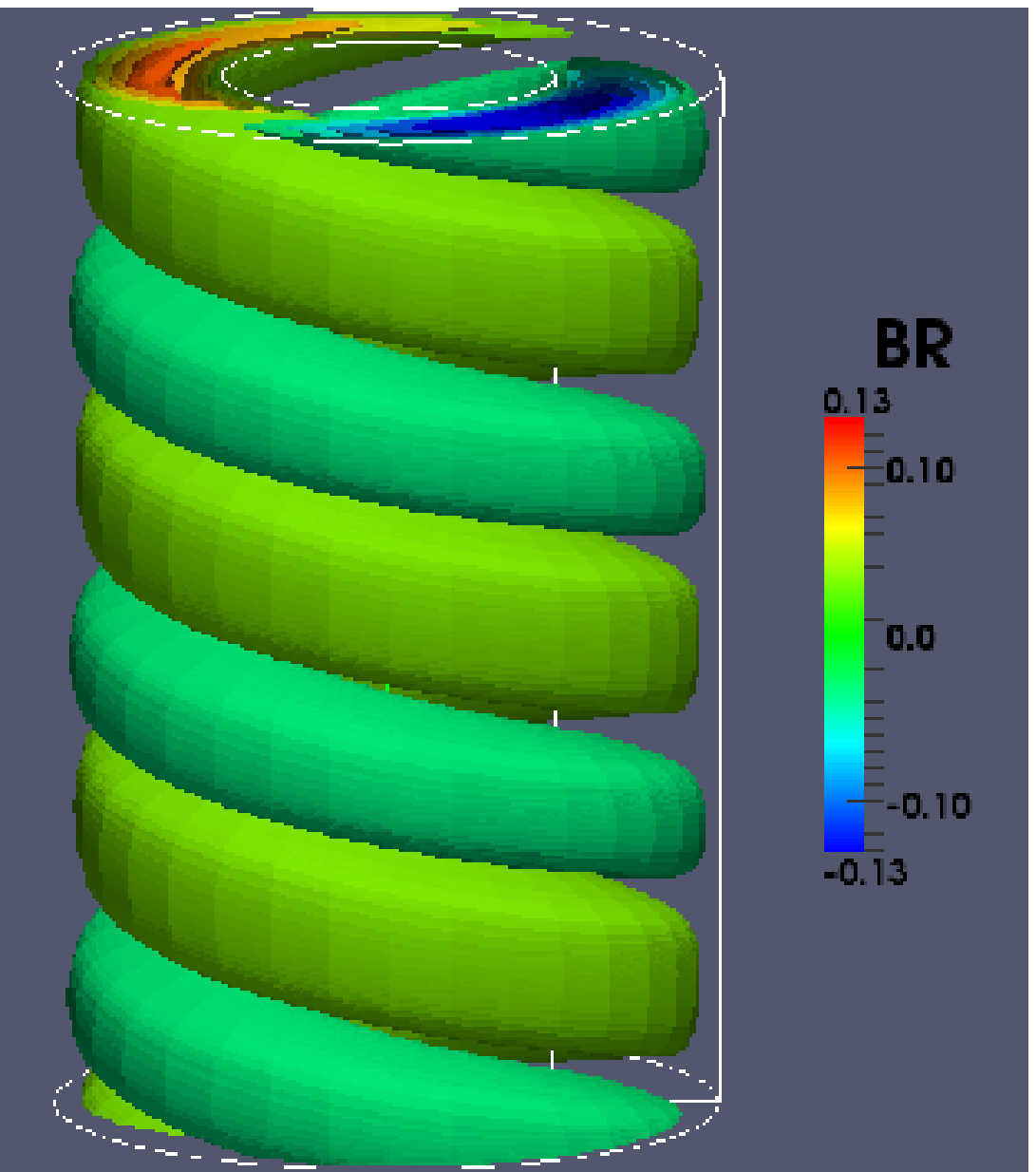,width=8.5cm,height=5cm}
\psfig{figure=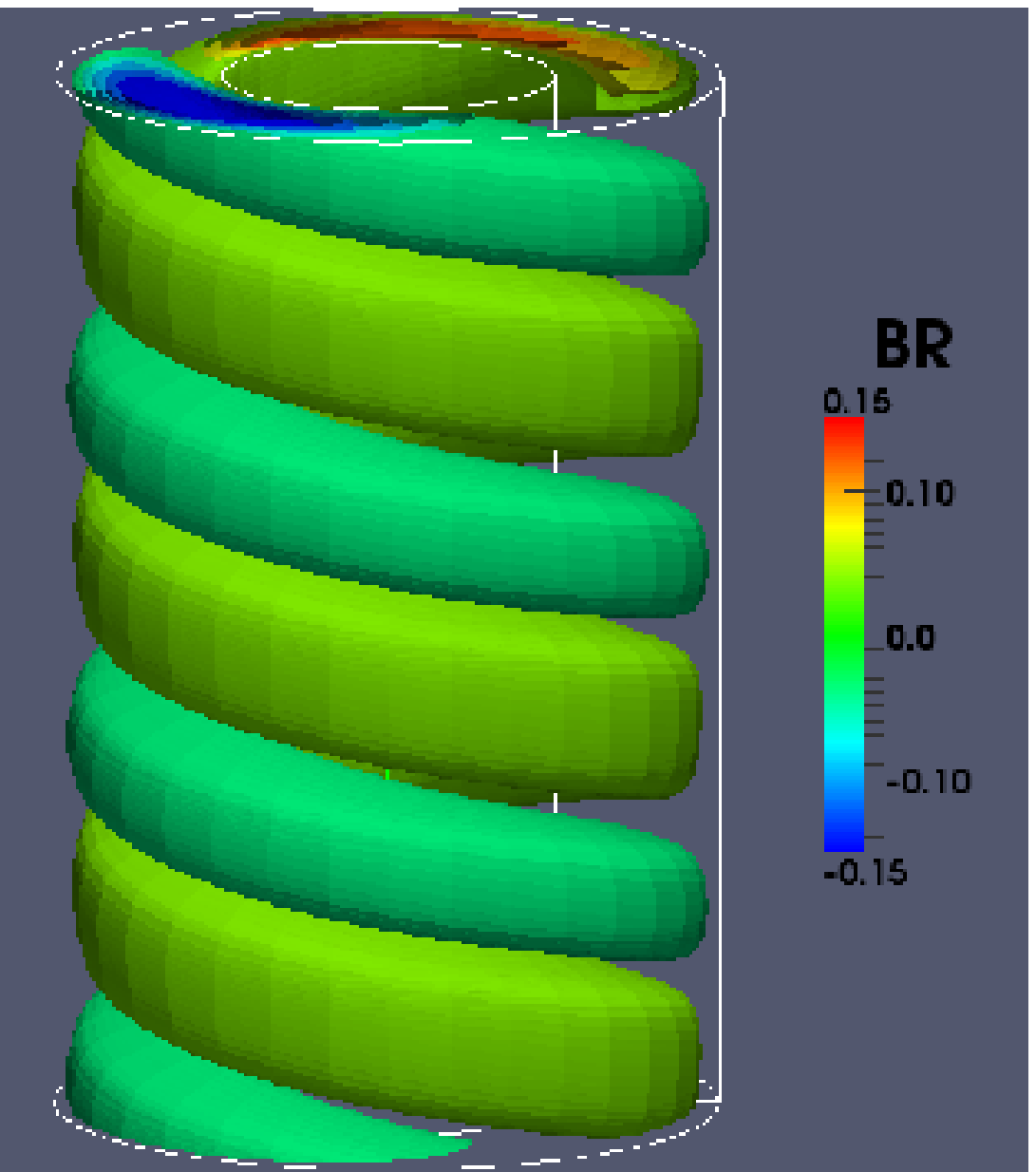,width=8.5cm,height=5cm}
}
\vbox{
\psfig{figure=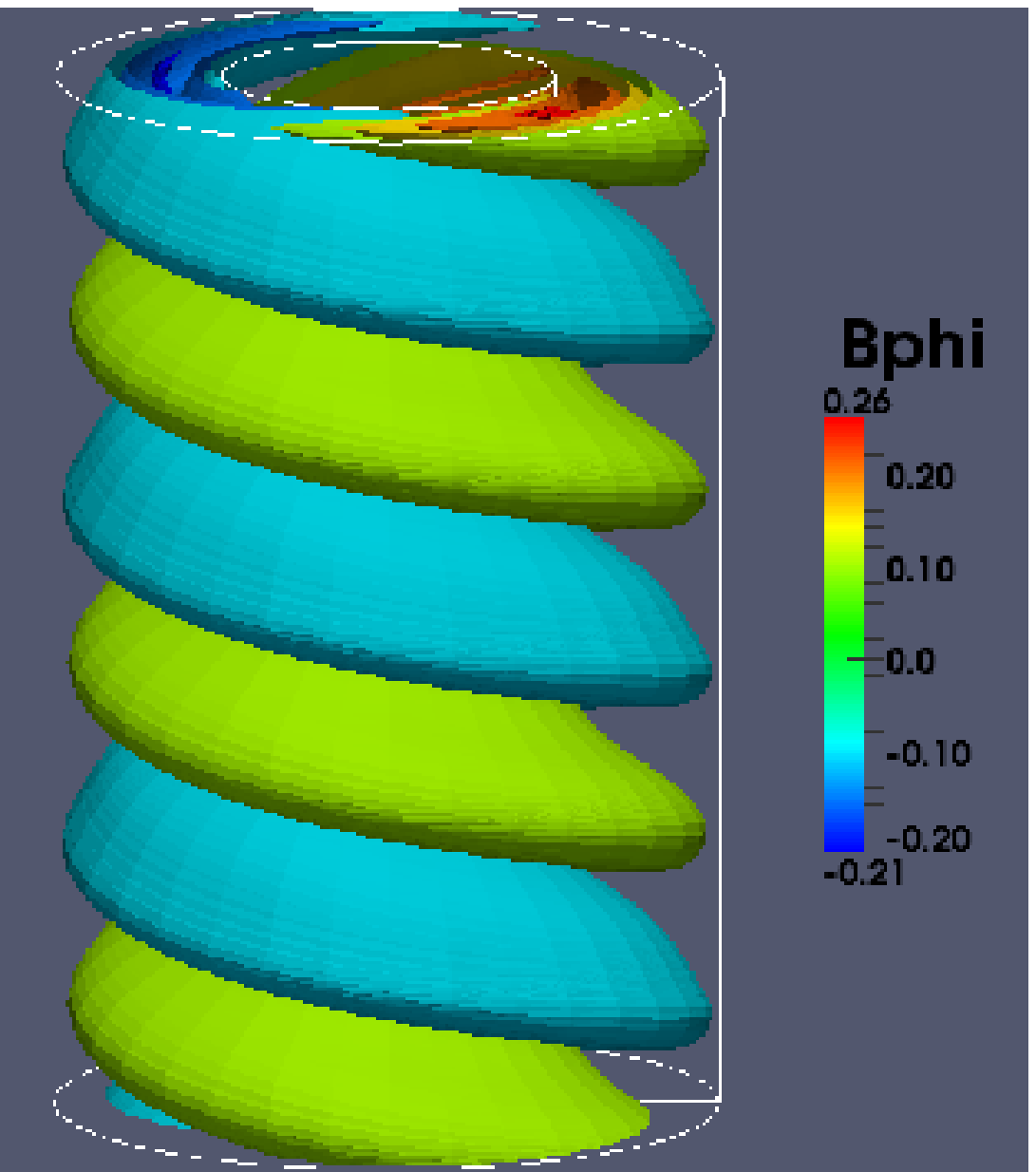,width=8.5cm,height=5cm}
\psfig{figure=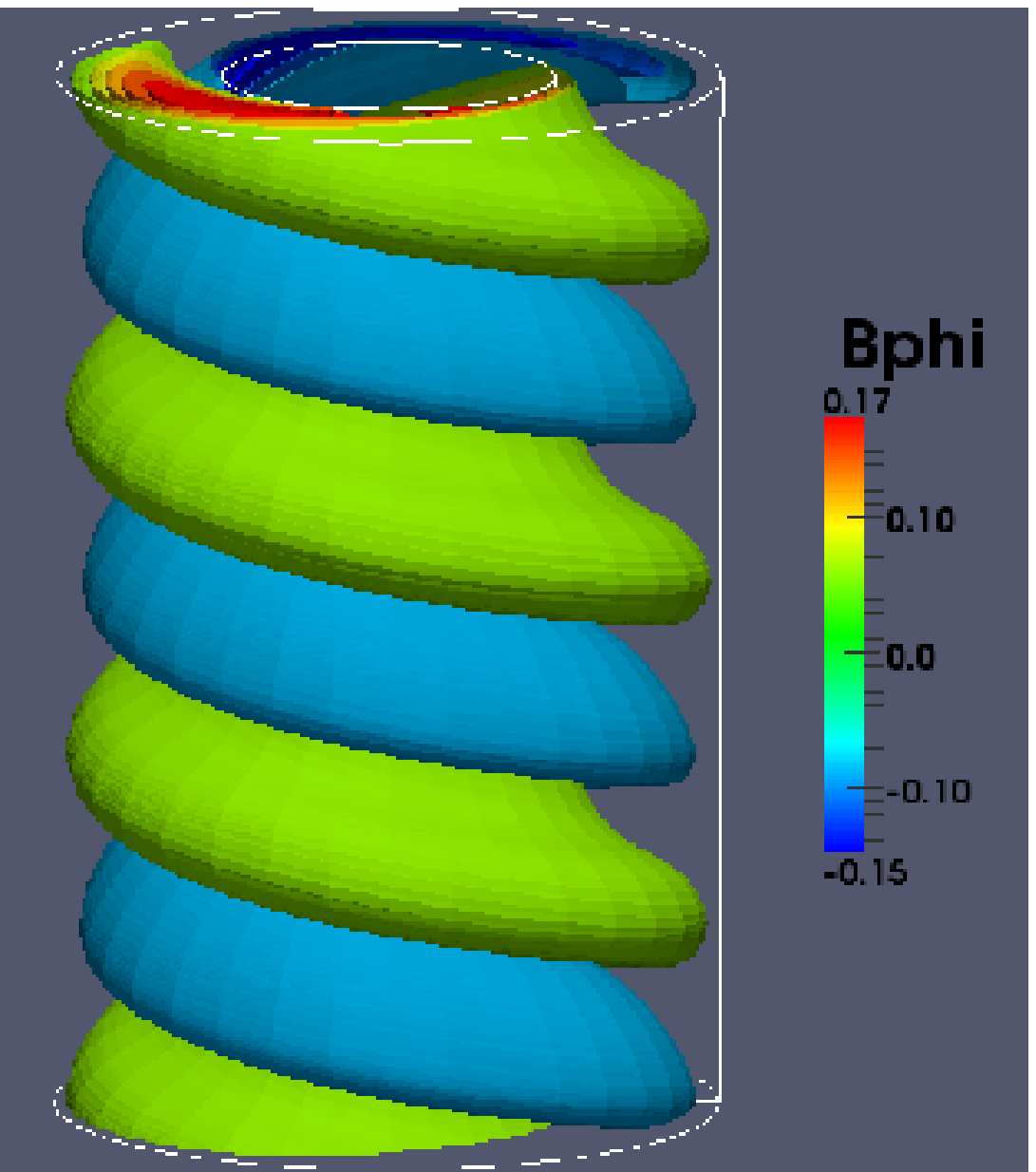,width=8.5cm,height=5cm}}
}
\caption{\label{f9} The components (left: radial component,  right: azimuthal
component) of the  magnetic pattern in the  AMRI domain (fast rotation) for $\beta=10$.
Top: $\rm Re=150,\ \  Ha=50$ ({ minimum}); bottom: $\rm Re=200, Ha=80$. The fields are normalized with $B_{\rm in}$.
 $\mu_B=1$, $\rm Pm=1$. }
\end{figure*}

Figure \ref{f9} concerns the AMRI domain. The value of $\beta=10$ is fixed,
but the location in the instability diagram differs slightly. The top row
shows the AMRI just for the minimum in Fig. \ref{f4},  while the
bottom row shows higher parameter values. In both cases though we see the
expected $m=1$ left spirals, in agreement with the linear results.

The simulations lead to a further basic result. By considering the maximum
values of the radial and azimuthal components, a distinct anticorrelation
becomes visible. The azimuthal component has its maximum where the radial
component has its minimum. The azimuthal average of $b_R b_\phi$,
is therefore negative. The magnetically driven angular momentum transport is
thus outward in both cases.
\begin{figure*}[htb]
\mbox{
\vbox{
\psfig{figure=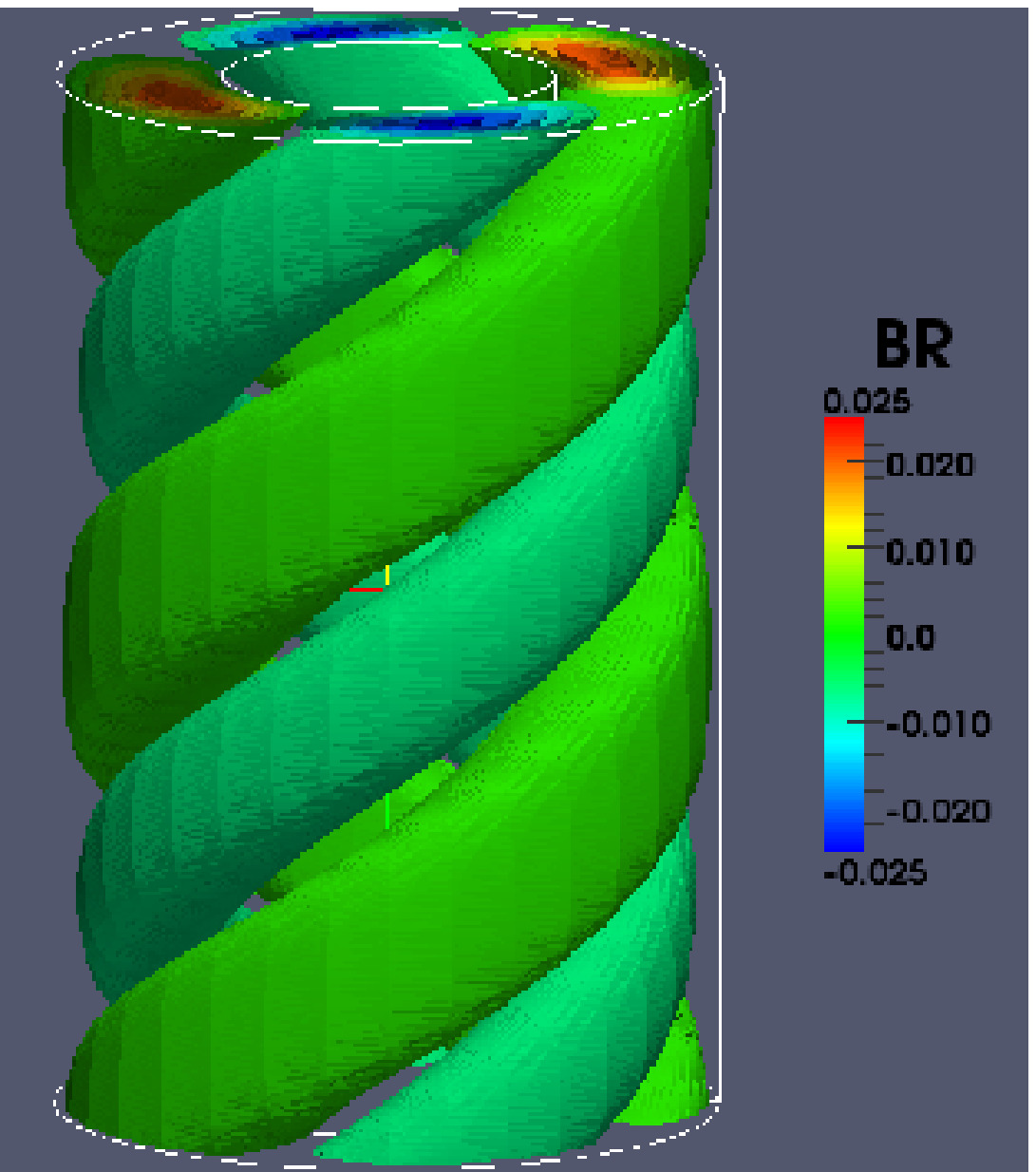,width=8.5cm,height=5cm}
\psfig{figure=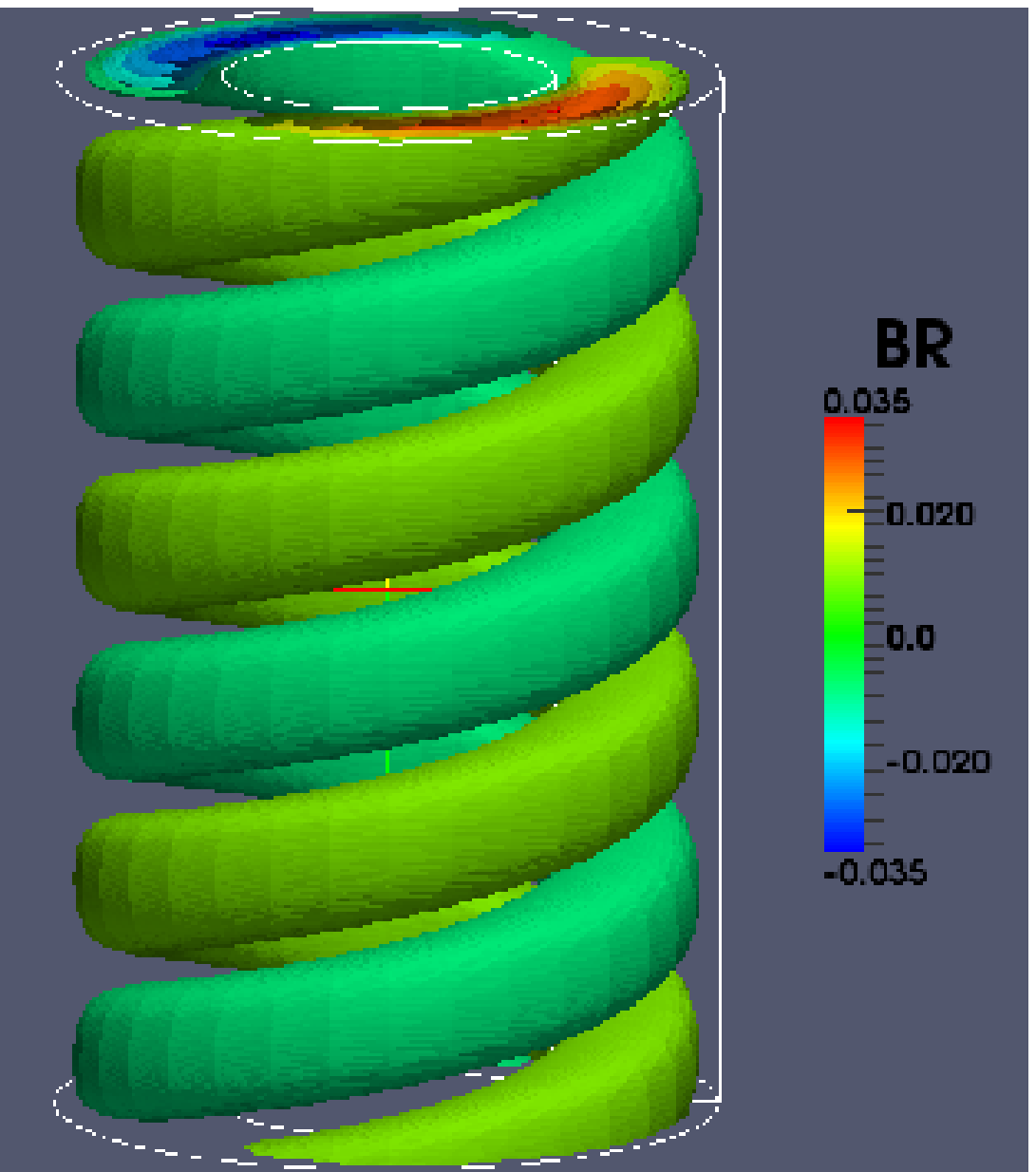,width=8.5cm,height=5cm}
}
\vbox{
\psfig{figure=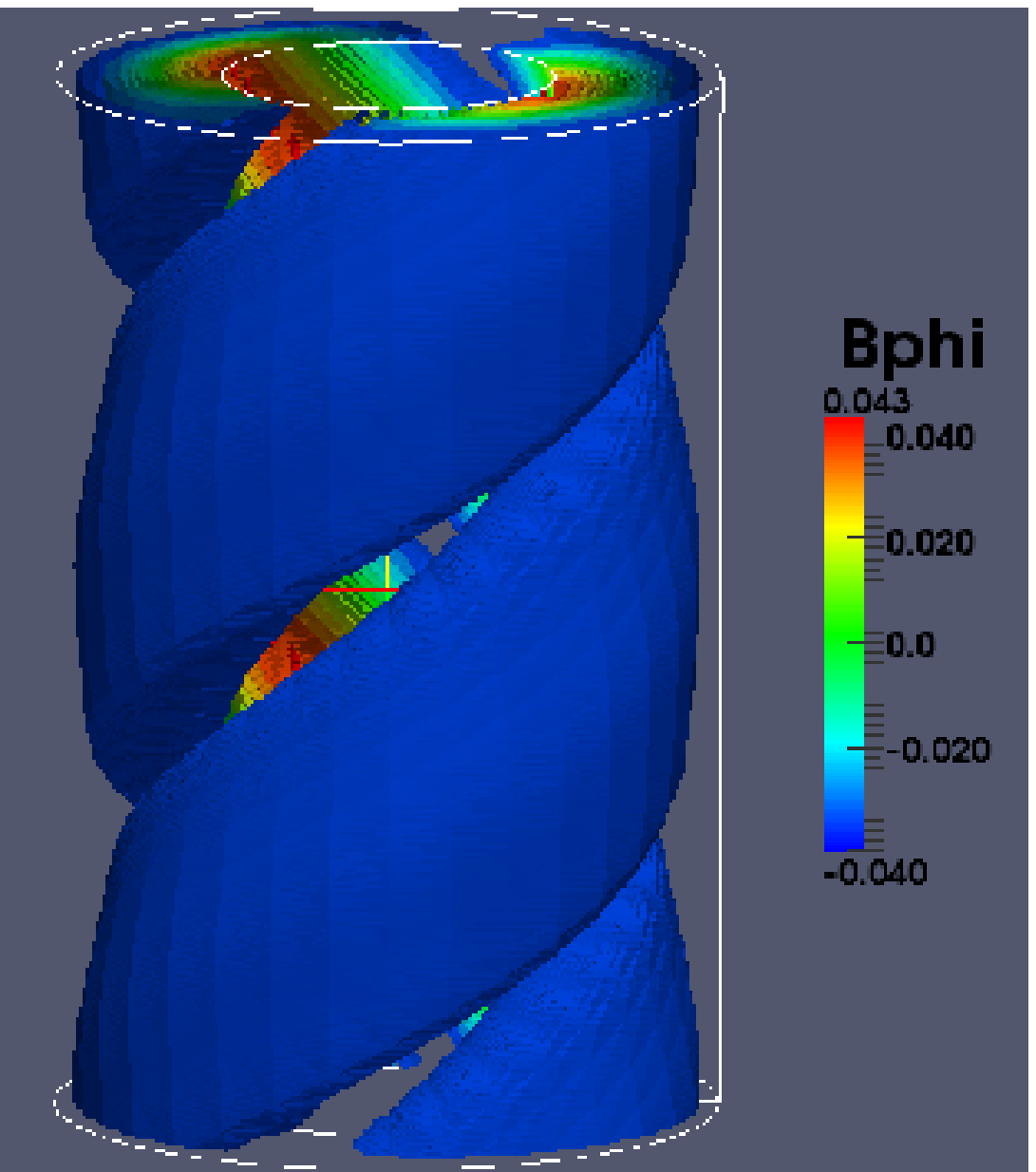,width=8.5cm,height=5cm}
\psfig{figure=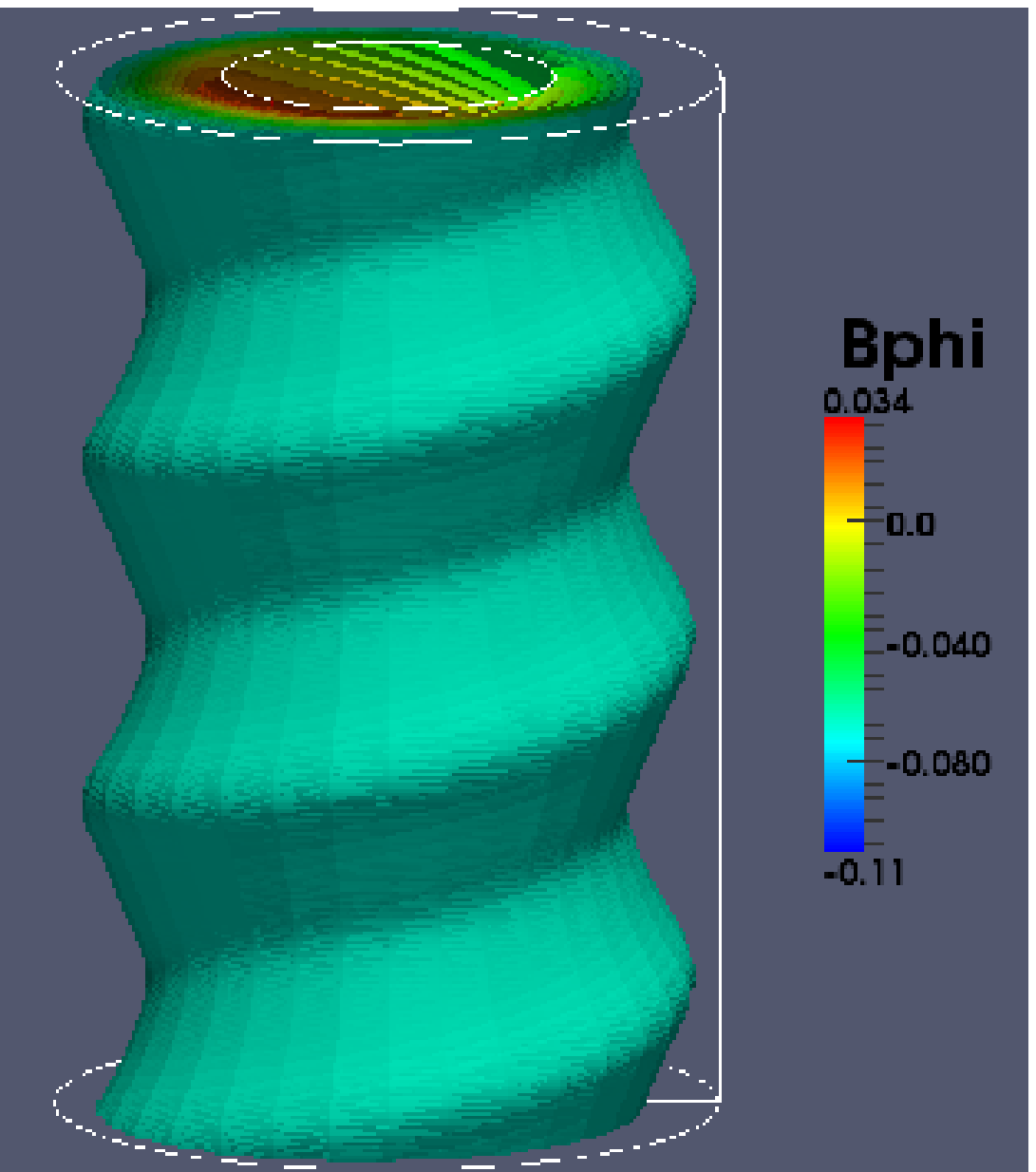,width=8.5cm,height=5cm}}
}
\caption{\label{f10} Components of the  magnetic pattern   (left: radial 
component,  right: azimuthal  component) for  TI (slow rotation).
$\rm Re=30$, $\rm Ha=130$. Top:  $\beta=2$,  Bottom: $\beta=10$. The fields are normalized with $B_{\rm in}$. $\mu_B=1
$, $\rm Pm=1$.}
\end{figure*}

Unlike the AMRI, the TI yields right-handed spirals (Fig. \ref{f10}). The
pattern in the bottom row ($\beta=10$) has an azimuthal wavenumber $m=1$,
in accordance with the instability map Fig. \ref{f4}. The top row, however,
represents a pattern with $m=2$, which also exists in the nonlinear 
regime as predicted by the bottom plot of Fig. \ref{f4}. In this case there
is no clear correlation between the radial and azimuthal components of the
field perturbations.

The nonlinear simulations, of course, do also provide the amplitudes of the fields in the resulting magnetic pattern. Here  we only  note the overall result that  the AMRI produces  much higher field strengths than the TI.  One might speculate that the AMRI  exists due to the  differential rotation which is always able to  induce strong fields but a detailed study of the energy aspects of the magnetic instabilities is  out of the scope of the present paper.

%%%%%%%%%%%%%%%%%%%%%%%%%%%%%%%%%%%%%%%%%%%%%%%%%%%%%%%%%%%%%%%%%%%%%%%%%%%%%%%
\end{document}